\documentstyle[12pt,epsf]{article}
\begin{document}
\baselineskip 16pt plus 2pt minus 2pt

\newcommand{\beq}{\begin{equation}}
\newcommand{\eeq}{\end{equation}}
\newcommand{\beqa}{\begin{eqnarray}}
\newcommand{\eeqa}{\end{eqnarray}}
\newcommand{\dida}[1]{/ \!\!\! #1}
\renewcommand{\Im}{\mbox{\sl{Im}}}
\renewcommand{\Re}{\mbox{\sl{Re}}}
\newcommand{\PRD}[3]{{Phys.~Rev.} \textbf{D#1}, #2 ({19#3})}
\newcommand{\PLB}[3]{{Phys.~Lett.} \textbf{B#1}, #2 ({19#3})}
\newcommand{\PRL}[3]{{Phys.~Rev.~Lett.} \textbf{#1}, #2 ({19#3})}
\newcommand{\NPB}[3]{{Nucl.~Phys.} \textbf{B#1}, #2 ({19#3})}
\newcommand{\NPA}[3]{{Nucl.~Phys.} \textbf{A#1}, #2 ({19#3})}
\def\simge{\hspace*{0.2em}\raisebox{0.5ex}{$>$}
     \hspace{-0.8em}\raisebox{-0.3em}{$\sim$}\hspace*{0.2em}}
\def\simle{\hspace*{0.2em}\raisebox{0.5ex}{$<$}
     \hspace{-0.8em}\raisebox{-0.3em}{$\sim$}\hspace*{0.2em}}

\begin{titlepage}


\hfill{TRI-PP-00-11}

\vspace{1.0cm}

\begin{center}
{\large {\bf The Use of Nuclear 
$\beta$-decay as a Test of Bulk Neutrinos in Extra Dimensions}}

\vspace{1.2cm}

G. C.  McLaughlin$^{\ast}$\footnote{email: gail@triumf.ca} and
J. N. Ng$^{\ast ,\dagger}$\footnote{email: misery@triumf.ca}

\vspace{0.8cm}

$^{\ast}$TRIUMF, 4004 Wesbrook Mall, Vancouver, BC, Canada V6T 2A3\\
$^{\dagger}$National Center for Theoretical Sciences, P.O. Box 2-131,
Hsinchu, Taiwan 300, Taiwan\\[.4cm]
\end{center}

\vspace{1cm}

\begin{abstract}
Theories which include neutrinos in large extra dimensions can be constrained
by nuclear beta decay experiments.  
We examine universality of $\beta$ decay
strengths of Fermi transitions.  From this we find that the extra dimensional
Yukawa coupling for a higher dimensional scale of 10 TeV, and two extra
dimensions can be constrained to $y \simle 1$.  Kinematic implications
are also discussed.  In particular, an extra dimensional 
scenario will
produce a tritium decay endpoint spectrum with a different shape than
that for just one massive state. 

\end{abstract}  

\vspace{2cm}
\vfill
\end{titlepage}
\section{Introduction}

It is well known that consistent string theory can only be formulated in space
time higher than four dimensions, specifically 10 or 11 dimensions.
Until recently the predominant view was that since the extra dimensions
have not been detected, they
must be compactified in small volumes with radii of Planck size.
This was supported by the study of the heterotic string where the string scale
$M_S$ is related to the Planck scale $M_P$ via \cite{PG}
\beq
 M_S= M_{P}\sqrt{k\alpha_{G}} \
\eeq
where $\alpha_{G}$ denotes the unified gauge coupling constant and $k$ is an
integer of order
unity which labels the level of the Kac-Moody algebra.  Here the
string scale is close to the Planck scale and this renders string effects unobservable.
Recently, Witten \cite{EW} observed
that in nonperturbative string theory $M_S$ can be much lower. Lykken \cite{LY} and
Antoniadis \cite{Ant} independently suggested the possibility that $M_S$
can be as low as a few TeV.  In this case, the
fundamental string scale is not constrained theoretically as above but
instead is constrained phenomenologically.
This
gains support from breakthroughs in understanding D-brane constructions
in string theory \cite{Pol}. In the D-brane scenario
the number of extra dimensions scanned by the Standard Model (SM) particles and
the graviton can be very different.

Buoyed by these developments intense efforts are made to understand the main features of low
scale string theories. A popular construction is to assume a factorizable geometry.
The gravitational field
is taken to propagate in the full 10 or 11-
dimensional bulk volume whereas the SM particles are localized in the 3-brane and
hence are not sensitive to
the extra dimensions \cite{EW, LY, AD, AAD}.  The crucial observation \cite{AD} is
that the weakness of gravity in four dimension is due to its spreading in the
$n$ extra dimensions
and Eq.(1) is replaced by
\beq
\label{fun}
 M^2_{P}=M^{n+2}_{\ast}V_n \;
\eeq
where $V_n$ is the volume of the compactified extra space and $M_{\ast}$ is the higher dimensional Planck
scale. The two scales
 $M_S$ and $M_{\star}$ are related but the exact relation will not be needed in this paper.
It is simplest to consider toroidal compactification, for which the volume is
\beq
 V_n=(2\pi)^{n}R_1R_2R_3\ldots R_n \;
\eeq
and $R_i(i=1,2,\ldots n)$ are the radii of the extra dimensions. For symmetric compactification one takes all
the radii to be equal and denoted it by $R$.

This formulation has dramatic effects that lead to a new
perspective on the mass hierarchy problem \cite{AD} and the gauge coupling
constants unification \cite{KD}.
Furthermore, these theories predict a deviation from the Newtonian $1/r$ law
of gravitational potential at submillimeter range. The fact that this law is well
tested above a distance of 1mm
leads to the limit that $n \geq 2$ for $M_{\ast}= 1 \, \mathrm TeV$, for the symmetric case.
New experiments currently underway
will probe smaller distance scales \cite{Lon}.
Besides the direct search for a deviation of the Newtonian
gravitational law the above scenario can also be indirectly
constrained by astrophysical considerations. Although
somewhat model dependent these considerations suggest
that $n \geq 2$ and $M_{\ast}\geq 30$ TeV \cite{CP}.

A different scenario to explain the weak gravitational force is proposed by Randall and
Sundrum \cite{RS}.  This mechanism
does not require large extra dimensions but instead invokes a non-factorizable geometry.
Here the extra dimension has the effect of modifying the space-time metric.
Experiments that probe gravity at small distances are
 then not interpreted in terms of the scaling law of Eq.~\ref{fun}.
The phenomenology of the graviton and its Kaluza-Klein (KK) excitations is different from the
case of factorizable geometry and has been  studied in Ref.\cite {HR}. Given the
differences of the these two
scenarios and their extensions, and the richness of the
phenomenology for each case it is clear that probes other than the
graviton and its KK spectrum will be important additions to the study of higher
dimensional physics.

It has been suggested recently \cite {ADDM,DDG}  that an extra right-handed
neutrino that is a standard model 
gauge singlet may also be a probe of the extra dimensions.
When a field is allowed to extend into the compactified extra dimensions it has
associated with it Kaluza-Klein excitations which can have detectable
effects in low energy experiments.  Neutrinos propagating
in the extra dimension will be referred to as the bulk
neutrinos. The couplings of the bulk neutrino states with the
active neutrino will give rise to a
small neutrino mass for the active neutrino due to the spreading
in  the bulk volume.
Some effects of these neutrinos in various low energy 
observables have been suggested \cite {Many}.
A solution to the solar neutrino problem in terms of matter enhanced flavor
transformation of $\nu_{eL}$ into bulk neutrino
states was studied in \cite{DS}. Bulk neutrinos can induce energy loss in stars
\cite {MN} and the possible connection to the supernova collapse phase is studied in
\cite{GF}. 

In this paper we concentrate on the constraints on the extra dimensional scenarios
from universality tests of nuclear $\beta$-decays. Such studies have been
used to
provide a strong basis for the SM and more recently their usage is mainly
in the test of 
unitarity of the quark mixings. 
We point out that the difference in  Q-values for the nuclei which are
very well measured can be used to constrain the
properties of the bulk neutrinos such as the
Yukawa couplings, the higher dimensional scale, $M_*$, and the
compactification radius R.
These are very general properties of neutrinos in extra dimensions and one does
not need to employ the scaling relation between the Planck scale and
$M_*$ as given in Eq.(\ref{fun}). To illustrate the effects
of the bulk 
neutrinos we give 
the Kurie plots of two nuclei and the nuclear recoil spectrum of
a Fermi transition.  The simplest model of
bulk neutrinos given below may allow a signal to be detected
in the $^3 {\rm H}$ beta decay spectra.  Other models which
modify the tritium beta decay endpoint are discussed in \cite{other}.  
However, in searches
for a small admixture of a more massive neutrino, the
features of the extra dimensional model are not within
currently detectable limits.  However more realistic models with larger
mixings with the active neutrinos may be constructed which can
be probed by these experiments as well.
 We concentrate only on one family of bulk neutrino
and the active $\nu_{eL}$. However, it is straightforward to extend the model
to include three bulk neutrinos; one
for each family.  This implies further model assumptions and will take us into the
realm of models for fermion
families in extra dimensions which is beyond the scope of this study.

This paper is organized as follow. In section \ref{sec:model}
we briefly outline the model of
bulk neutrinos that we use and the mixings with $\nu_{eL}$.
This serves the purpose of fixing of our notation. Sec. \ref{sec:univ}
details the study of universality using the data
from Q-values of several $\beta$-decays. Next
we present the impact of bulk neutrinos on the beta spectrum, and look
at the recoil momentum for Fermi transitions.
We give our conclusion in Sec \ref{sec:concl}.

\section {A Simple Model of Bulk Neutrinos}
\label{sec:model}

The basic idea of producing neutrino masses using extra dimensions is
given in Ref. \cite{DS}. In the interest of being self contained and to
establish our notation we give some of the construction of a simple
model of bulk neutrinos. The simplest model
assumes that the fermions charged under the standard model gauge group
as well as the gauge bosons and
the Higgs boson are localized on a 3-brane embedded in the bulk of 
larger dimensions.  If we assume only
one extra dimension then space-time is labelled by $(x^{\mu},z)$ 
where $\mu =0,1,2,3$.
The extra coordinate
$z$ is assumed to compactify into a circle of radius R. The brane scenario 
stipulates that the
left-handed SM lepton
doublet $L=(\nu_{eL}, e_L)$ is given by $L(x^{\mu},z=0)$. The fields 
$\nu_{eL}$ and $e_L$ are
Weyl spinors.
Next we assume that a SM singlet fermion, denoted by
$\nu(x,z)$ exists and it
propagates in the full five dimensional bulk. Its right- and
left-handed projections are
labelled by $\nu_R$ and $\nu_L$ respectively. It is distinguished from the
SM neutrino by not carrying a flavor index. In this simple model gravity
is weak because there are additional dimensions in which it propagates.

In five dimensions there are five gamma matrices $\Gamma^{\mu}$ and $\Gamma^5$ where $\mu =0,1,2,3$.
A convenient
representation of the Clifford algebra in five dimensions is to choose
$\Gamma^{\mu}=\gamma^{\mu}$ and $\Gamma^5=i\gamma^5$ with the usual $\gamma^{\mu}$ of Minkowski
space-time.
The effective Lagrangian for generating a neutrino mass is given by
\beq
\label{lm}
{\cal L}=\int_0^{2\pi R} dz \bar{\nu}\left( i\gamma^{\mu}\partial_{\mu}+i \Gamma_5
\partial_z \right)\nu+
y_{\ast}\int_0^{2\pi R}dz \delta (z) \bar{L} H \nu_R + h.c.    \;
\eeq
where $y_\ast$ is the dimensionful Yukawa coupling and $H$ is the Higgs doublet and it is 
related to the dimensionless coupling $y$ via
\beq
\label{yus}
y_{\ast}=\frac{y}{M_{\ast}^{n/2}} \,
\eeq
and $n=1$ for one extra dimension.

For simplicity we have neglected a higher
dimensional bare
Dirac mass term. This can be naturally implemented under ${\cal Z}_2$ orbifold
 compactification \cite {DDG}. Following the Kaluza-Klein ansatz we Fourier
 expand $\nu_R$ and $\nu_L$ as follows:
\beqa
\nu_R(x,z)&=&\frac{1}{\sqrt{2\pi R}}\sum_{k=-\infty}^{\infty}\nu_{kR}{\mathrm{exp}}\left(\frac{ikz}{R}\right)\\
\nu_L(x,z)&=&\frac{1}{\sqrt{2\pi R}}\sum_{k=-\infty}^{\infty}\nu_{kL}{\mathrm{exp}}\left(\frac{ikz}{R}\right) \,
\eeqa
Substituting this into Eq.(\ref{lm}) and integrating over $z$ yields the effective Lagrangian in four dimensions:
\beq
\label{leff}
{\cal L} = \sum_{k=-\infty}^{\infty}\left[ \bar{\nu}_{kL}i\gamma^{\mu}\partial_{\mu}\nu_{kL} +\bar{\nu}_{kR}
i\gamma^{\mu}\partial_{\mu}\nu_{kR} \right]+ \sum_{k=-\infty}^{\infty}m_{k}\bar{\nu}_{kL}\nu_{kR} +
\frac{y_\ast}{\sqrt{2\pi R}}\sum_{k=-\infty}^{\infty}\bar{L}H \nu_{kR} + h.c.
\eeq
where
\beq 
\label{kkm}
m_k=\frac{k}{R}
\eeq
 is the mass of the $k-\mathrm th$ KK tower state.
The mass splitting between each adjacent
tower state is $1/R$. As seen in Eq.(\ref {leff}) the coupling between
the KK-tower neutrino states and the active $\nu_{eL}$ given
by the Yukawa term. After spontaneous electroweak symmetry breaking 
a Dirac mass term is generated for
the active $\nu_{eL}$ and is given by
\beq
\label{md}
m_D=\frac{yv}{\sqrt{4 \pi RM_{\ast}}}
\eeq
where $v= 247$ GeV. This mass is suppressed by a bulk volume factor as 
first noticed in Ref. \cite {ADDM} and \cite {DDG}.

	Rewriting all the mass terms in Eq.(
\ref{leff}) we have
\beq
\label{nud}
m_D\bar{\nu}_{eL}\nu_{0R} + m_D\sum_{k=1}^{\infty} \bar{\nu}_{eL}
(\nu_{kR}+ \nu_{-kR})
+\sum_{k=1}^{\infty}m_k(\bar{\nu}_{kL}\nu_{kR}-\bar{\nu}_{-kL}\nu_{-kR}) + h.c. \,
\eeq
We find it useful to define the following orthogonal states
\beq
\nu^{'}_{kR} =\frac{1}{\sqrt {2}}(\nu_{kR}+\nu_{-kR}) \hspace{2.cm}
\nu^{''}_{kR}= \frac{1}{\sqrt {2}}(\nu_{kR}-\nu_{-kR}) \;
\eeq
and
\beq
\nu^{'}_{kL}=\frac{1}{\sqrt {2}}(\nu_{kL}-\nu_{-kL}) \hspace{2.cm}
\nu^{''}_{kL}=\frac{1}{\sqrt {2}}(\nu_{kL}+\nu_{-kL}) \:
\eeq
which can be used to cast Eq.\ref{nud} into the form
\beq
\label{nudd}
m_D\bar{\nu}_{eL}\nu_{0R} + \sqrt {2}m_D\sum_{k=1}^{\infty}\bar{\nu}_{eL}\nu_{kR}^{'}
+
\sum_{k=1}^{\infty}m_k(\bar{\nu}^{'}_{kL}\nu^{'}_{kR}+\bar{\nu}^{''}_{kL}\nu^{''}_{kR}) +h.c. \,
\eeq
The states with double prime superscripts have no low energy interactions and will be ignored. The mass
terms can now be written in the familiar form of $\bar{\nu}_L M\nu_R$ in the bases of $\nu_L=
(\nu_{eL},\nu_{1L}^\prime,\nu_{2L}^\prime,\ldots)$ and $\nu_R=(\nu_{0R}^\prime,\nu_{1R}^\prime,
\nu_{2R}^\prime,\ldots)$. The mass matrix $M$ for $k+1$ states looks as follows
\beqa
M=\left( \begin{array}{ccccc}
m_D&\sqrt {2}m_D &\sqrt {2}m_D &\dots &\sqrt {2}m_D  \\
0  &\frac{1}{R}&0 &\dots  & 0\\
0  & 0     &\frac{2}{R} &\dots  &0   \\
\dots &\dots &\dots &\dots &\dots\\
0 &  0  & 0 & 0 &\frac{k}{R}
\end{array} \right)
\eeqa

To find the left mass eigenstates, we consider
the matrix
$MM^{\dagger}$. Explicitly,
\beqa
\label{M2}
MM^{\dagger}=\frac{1}{R^2}\left(\begin{array}{ccccc}
(k+\frac{1}{2})\zeta^2 & \zeta & 2\zeta & \dots & k\zeta \\
\zeta & 1& 0 &\dots & 0 \\
2\zeta & 0 & 4 & \dots & 0\\
\dots & \dots & \dots &\dots & \dots \\
k\zeta & 0 & 0 & 0 & k^2
\end{array} \right)
\eeqa
where $\zeta=\sqrt{2}m_DR$.
The mass eigenvalues $\lambda $ are given by the characteristic equation
of
Eq.(\ref{M2}) which is det$[MM^{\dagger} -\lambda^2]$. After some
algebra it reduces to the transcendental equation \cite {DDG}:
\beq
\label{cheq}
\pi\zeta^2\cot(\pi \lambda R)=2\lambda R \;
\eeq 
The solution of this equation yields the lowest mixed  mass eigenstate
$\simeq m_D$. For small $\zeta$ this is also the maximally mixed state. 
It is reasonable to assume that 
 experiments that directly measures the neutrino mass will be probing
this state.
The current experimental limit on the mass of $\nu_e$ is $m_{\nu_e}
< 2.5 \, \rm eV$  \cite {nue}.
When combined with this, Eq. (\ref{md}) gives the following constraint
\beq
\label{c1}
y \simle 3.2\times 10^{-3}\left( \frac{M_*}{10 \rm TeV}
\right)^{\frac{1}{2}}
\left( \frac{R}{1 \rm mm} \right)^{\frac{1}{2}}\left( \frac{m_D}{1 \rm eV}
\right) \;
\eeq
The diagonalization of Eq. (\ref{M2}) leads to the mixing 
of the $k$-th state to
the lowest mass eigenstate 
$\nu_{0L}^m$ given by \cite{DS}
\beq
\label{mixe}
\tan {2\theta_k}= \frac{2k\zeta}{k^2-(k+\frac{1}{2})\zeta^2} \;
\eeq
This equation diagonalizes the submatrix between the lowest state
$\nu_{0L}$ and the state $\nu_{kL}$. Where $\zeta$ is small, this
is a good approximation to the exact mixing angle, 
obtainable by complete diagonalization of the matrix MM$^\dagger$.
In this limit, $\zeta$ can be used as an
expansion parameter in the
diagonalization process. For example, for $M_*
= 10 \, {\rm TeV}$,
$y = 10^{-3}$ and
$R\sim 10^{-7} \,{\mathrm mm}$ we find $\zeta\sim 10^{-1}$.
Hence, for small $\zeta$  we can see from above 
that the mixing of the KK states
with the lowest mass neutrino eigenstate becomes progressively smaller
 as the value of $k$ increases.
Furthermore, these KK states have no direct gauge interactions with other
SM particles and their
presence can only be probed through their mixing with the state $\nu_{0L}^m$.

In the small $\zeta$ limit, the vacuum oscillation survival probability for 
electron neutrinos is a particularly simple expression,
\begin{equation}
P(\nu_e \rightarrow \nu_e) \approx 1 - {\pi^2 \zeta^2 \over 3} 
+ 2 \sum_k {\zeta^2 \over k^2}  \left[ 1 - 2 \sin^2 
\left( \delta m_{0k}^2 t \over 2 E_\nu \right) \right] + {\cal O}(\zeta^4)
\end{equation}
where $\delta m_{0k}^2 \approx m_D^2 - m_k^2$.  The data from a reactor neutrino
experiment can be used to find a limit on the Yukawa coupling.  CHOOZ \cite{chooz}
quotes a limit on two neutrino mixing of $ \sin^2 2 \theta < 10\%$ for 
$\delta m^2 > 10^{-3} {\rm eV}^2$.
This translates into a limit on the extra dimensional parameters of
\begin{equation}
\label{eq:mixing}
y  < 4.4 \times 10^{-8} \left( {1 {\rm mm} \over R} \right)^{1/2} 
\left( { M_* \over 10 \, {\rm TeV} } \right)^{1 \over 2}.
\end{equation}
If combined with the limit from the neutrino mass, Eq \ref{c1},
we find that the least
 stringent limit on $y$ occurs at $R = 7 \times 10^{-6} {\rm mm}$.

	However, Eq.(\ref{mixe}) reveals a different phenomenon if
$\zeta$ is large; i.e. $\zeta \gg 1 $. Now it is the higher KK states
with $k \approx \zeta$ that will have the largest mixing with
the the lowest eigenstate; whereas states with lower $k$ values will have
small mixing of order $\frac{1}{\zeta}$. 
In order for this to happen in $n=1$ case we require 
 $R \geq 10^{-3}$ mm. Obviously this
 cannot happen with $\zeta \leq 1$ and a small compactification radius. 
When relatively large values of $R$ are considered, 
masses of these KK states can be
in the $ \sim 1$ eV range. 

	The above considerations can be generalized to higher extra
dimensions. For $n=2$ and different compactification radii $R_1$ and $R_2$ 
the scaling equation for $m_D$ becomes
\beq
\label{md2}
m_D=\frac{yv}{2\pi M_*\sqrt{2R_1R_2}} \;
\eeq
and yields the following constraint the on the Yukawa coupling
\beq
\label{y2}
y\simle 1.8\times 10^{6}\left( \frac{M_*}{10 \rm TeV}
\right)\left(\frac{R_1R_2}{1\rm mm^2} \right)^{\frac{1}{2}} \left( \frac{m_D}{1 \rm eV}
\right) \;
\eeq
For $R_1=R_2$ and a comparison with Eq.(\ref{c1}) shows that this is
a much less stringent constraint on the parameters of the theory compared
to the case of only one extra dimension. This is expected since the bulk
neutrino state has a larger volume in which to spread and the Yukawa coupling 
is inversely proportional to the square root of this volume.
An interesting
case occurs for
 asymmetric compactification where one radius is much smaller than
 the other. Take for example
 $R_2 = 10^{-11} \rm mm$ and $R_1= 0.1 \, \rm mm$
then Eq.(\ref{y2}) approaches the case of $n=1$ and gives a limit on $y$
similar to that of Eq.(\ref{c1}). 

        The masses of the KK tower states are generalized to 
\beq
\label{n2m}
m_{k,l}=\sqrt{ \left( \frac{k^2}{R_1^2} + \frac{l^2}{R_2^2} \right) } \;
\eeq
where $k$ and $l$ denote the KK level corresponding to radii $R_1$ and $R_2$ respectively. It
is seen that the smaller compactification radius $R_2$ will give rise to higher mass KK state
in the asymmetric compactification scenario. For example if $R_2 =10^{-7}
\rm mm$ one can have
keV bulk neutrinos mixing into the active $\nu_e$ state that come from
the
small radius (i.e. 
for $k=0$ and small $l$ values). When both radii are large such neutrinos
come only from
the high KK values. Next we examine the mixings of these bulk states.

        For small mixing the electron neutrino state can written as \cite{DS}
\beq
\label{ne2}
\nu_e = {1 \over N} \left( \nu_0 +\zeta_1 \sum_{k=1}\frac{1}{k}\nu_{k,0} + \zeta_1 \sum_{k,l\geq 1}\frac
{1}{\sqrt{k^2+(\frac{R_1}{R_2})^2l^2}}\nu_{k,l} \right) \;
\eeq
where $\zeta_1=\sqrt{2}m_{D}R_1 <<1$ and $N$ is the normalization constant.  
It is seen that for the asymmetric case the 
KK states which come solely from 
the small radius may have larger masses than those
from the larger radius; moreover, they are accompanied by
correspondingly  smaller
mixings. We conclude that high mass states will have small mixing,
regardless of the
size of the compactification radii mainly due to the constraint on $m_D$
from the
direct limit on the electron neutrino mass.  Since the direct limits
on the mu and tau neutrino masses are much less strict, large mixing
may occur in these families for considerably smaller KK masses.
 
For two extra dimensions, the constant, $N$, depends on the cutoff scale 
$M_*$, as $N \sim 1 + \zeta^2 \log (M_* R)$, if $R_1 = R_2$ = R.  Limits
derived from universality of Fermi transitions which are presented in 
the next section do not depend on this normalization and therefore on
the cutoff procedure.  However, other calculations such as vacuum oscillation
survival probabilities will depend on the normalization.  In particular, this
would become important if one were to compare the theory with the 
reactor neutrino data.  In this case,  the constraint on $y$ and $M_*$ 
from universality could be used together with survival probability 
data to limit the size, $R$.   

        In the next section we study the constraint $\beta$ decay 
universality places on the parameters $y, M_*$ and $R$.

\section {Universality}
\label{sec:univ}

In this section we study the consequences for beta decay
of the model for bulk neutrinos outlined in section \ref{sec:model}.
Although bulk neutrinos do not have gauged interactions with SM particles,
the allowed Yukawa coupling enables them to generate mass for the active
neutrino. As seen in the last section the smallness of the neutrino mass
is due to the spreading of the state in the  higher dimensions. The
KK excitations are also seen to have mixing effects with the active
neutrino. In a given weak decay, the number of KK states that can be excited 
will depend on the energy released in the reaction.  If $ 1 /R << Q$, where 
$Q$ is the nuclear Q-value, then many states contribute.  For example, for 
n = 1, if $Q = \, 1 {\rm MeV}$ and $R = 10^{-7} \, {\rm mm}$ then
around 500 states contribute. In previous studies of such
classical weak interactions one concentrates on  one 
or two massive neutrino
states and their kinematic effect. In the present scenario, towers of
KK neutrinos ranging in mass from eV to a few MeV are involved in
a nuclear $\beta$ decay. They have to be taken into account in the study
of universality, electron spectra and nuclear recoil spectra. We shall
examine these issues separately to see if they
provide useful constraints on the Yukawa coupling, $y$ and the size R, of
the extra dimension.  We begin with universality.

The rate of beta decay from a single state in the parent nucleus to another state
in the daughter nucleus is given by,
\begin{equation}
\label{eq:rate}
\lambda = { \ln 2 \over ft} P(m_\nu)
\end{equation}
  The $ft$- value is
proportional to the inverse of the matrix element.
$P(m_\nu)$ is the phase space factor, which depends on the mass of the emitted
neutrino, $m_\nu$. We define the phase space factor as
\begin{equation}
\label{eq:phase1}
P(m_\nu) = { 1 \over m_e^5}
\int^{Q +m_e -m_\nu}_{m_e} F(Z,E_e) E_e p_e (Q + m_e - E_e)
[(Q +m_e -E_e)^2 - m_\nu^2]^{1/2} dE_e
\end{equation}
The  Q-value is the energy difference between the initial and final nuclear
states, while $E_e$, and $p_e$
are the electron energy and momentum respectively.
The Coulomb wave correction factor is denoted by $F(Z,E)$.  We use the
approximation of this factor found in \cite{ffn}.

For pure Fermi, $0^+ \rightarrow 0^+$ transitions and
zero mass electron neutrinos, the
corrected $ft$-values,
$Ft = ft( 1+ \delta_R) ( 1 - \delta_C)$  for all nuclei
should be the same under the
assumption of universality in the standard model.  Here,   $\delta_R$ is the nucleus
dependent part of the radiative correction and $\delta_C$ is the
isospin symmetry breaking correction.  These corrections contribute at
the 1\% level. Results of Ft-values for 10 nuclei are listed in
\cite{towner}.

The bulk neutrino scenario outlined in the previous section would cause an
apparent deviation from the constant $Ft$ values predicted by universality
in the standard model.  This is because the actual phase space factor in an extra
dimensional scenario is more complicated than Eq.(\ref{eq:phase1}).  In
general
$ft$-values are found by using Eq. \ref{eq:rate}, with a
measured rate and a calculated phase space factor.
The transition rate for the case of one dimension is given in the small
mixing limit, $\zeta <<1$, by
\begin{equation}
\lambda = {\ln2 \over ft_{xd}} \left[ \left( 1 - {\pi^2 \zeta^2 \over 6} \right)
P(m_\nu \approx 0)
+ \sum_{k = 1}^{k_{max}} P(m_{\nu_k} \approx {k \over R}) {\zeta^2 \over k^2} \right]
\end{equation}
The maximum number of neutrinos to be
summed over is determined by the maximum mass of a neutrino that can be released in the
beta decay, $k_{max} = Q R$. Transitions with a higher Q value will have more
neutrino states that can contribute to the decay.  Note that the $ft$ value in an extra
dimensional scenario will not necessarily take on the same value as given in the
standard model. (Although, the \lq true\rq\ $Ft$- values will be the
same for all $0^+ \rightarrow 0^+$ transitions.)
This will change the calculated value of the quark mixing angle
$V_{ud}$ as discussed below.
Because of this uncertainty, it is best to employ a normalization when looking for the
effects of bulk neutrinos.  In this example we will
compare the relative difference in beta decays from two different nuclei.

Assuming that bulk neutrinos exist in one extra dimension,
we can compare the $ft$-value for the extra dimensional scenario with the
apparent $ft$-values for the standard model for the nucleus, $A_1$, by taking a ratio,
\begin{equation}
{ft(A_1)_{xd} \over ft(A_1)_{SM,apparent}} \approx
\left( 1 - {\pi^2 \zeta^2 \over 6} \right) +
\sum_{k = 1}^{k_{max}} {P(A_1,m_{\nu_k}) \over P(A_1,0)} {\zeta^2 \over k^2},
\end{equation}
where we have used ${\lambda_{SM} /  \lambda_{xd}} = 1$.
Since in the extra dimensional scenario,
the $ft$-values for two nuclei are the same up to the isospin
and radiative corrections, the apparent Standard Model values will not be the same,
\begin{eqnarray}
{ft(A_2)_{SM, apparent} \over ft(A_1)_{SM, apparent}} \approx  1 && + \sum_{k = 1}^{k_{max} = Q_1 R} {P(A_1,m_{\nu_k}) \over P(A_1,0)} {\zeta^2 \over k^2} - \sum_{k = 1}^{k_{max} = Q_2 R} {P(A_2,m_{\nu_k}) \over P(A_2,0)} {\zeta^2 \over k^2} \nonumber \\
& & - [\delta_R(A_1) - \delta_R(A_2)] + [\delta_C(A_1) - \delta_C(A_2)].
\end{eqnarray}
The maximum value of the second two terms may be obtained from the experimental
uncertainty in the $Ft$-values from \cite{towner}.  The sums in the above expression
are very insensitive to lower bound on the sum, so it is only necessary to consider
the highest modes in the sum, where there are relatively large differences 
in the phase space factor ratio.  The mixing of the low energy KK states is not
relevant to universality considerations.  

The above formula is
valid for large $\zeta$, provided that ${\zeta^2 /  k^2} <<1$, if the
term $(\zeta^2 / k^2)$ is replaced by $ f(\zeta)^2 (\zeta^2 / k^2)$.
Here $f(\zeta)$ is a function which must be calculated for each $\zeta$
by explicit diagonalization of the mass matrix
given by Eq.(\ref{M2}).  This explicit diagonalization  
shows that the mixing of state with $k >> m_D R$ is
approximately  $\zeta^\prime / k = f(\zeta) \zeta /k$.  For $\zeta = 100$, 
$f(\zeta) = 0.007$.  For larger $\zeta$, $f(\zeta)$ decreases, while
for $\zeta <<1$, $f(\zeta) = 1$.  

We use $\beta^+$ decay
data from $^{14}{\rm O}$ and $^{54}{\rm Co}$ with a combined error of $0.22\%$.
These nuclei have Q values of 1.81 MeV and 7.22 MeV respectively.
This can be translated into a limit on the Yukawa coupling,$ y$ and the
size of the
extra dimension, R,
\begin{eqnarray}
\label{eq:limone}
0.0022 \simge 1.9 \times 10^{-7} {\rm MeV}^2
y^2 f(\zeta)^2 \left( {10 \, {\rm  TeV} \over M_*} \right)
\left( {1  \, {\rm  mm} \over R} \right) && \nonumber \\
\Biggl[
\sum_{k = 1}^{Q_1 R} \left( {P(A_1,m_{\nu_k}) \over P(A_1,0)}
-{P(A_2,m_{\nu_k}) \over
P(A_2,0)}  \right) { 1 \over m_{\nu_k}^2} 
- \sum_{k = Q_1R}^{Q_2 R} {P(A_2,m_{\nu_k}) \over P(A_2,0)} { 1 \over m_{\nu_k}^2} \Biggr]
\end{eqnarray}
Converting the sums in the above equation into integrals and integrating
 over all the available neutrino states produces a limit on the
Yukawa coupling.
As long as the small mixing limit applies, the 
limit on $y$ has little R dependence and remains fairly constant for example, 
at $y < 10^{-3}$
for a scale of  $M_* = 10 \, {\rm TeV}$.  The limit becomes
less strict for a larger high dimensional scale. 
The lack of R dependence can be seen when converting the sum in Eq.
(\ref{eq:limone})
to an integral, and making the change of variables $ dk \rightarrow R dm_\nu$.
In the case of large mixing, the limit is less strict, due to the factor
$f(\zeta)$.

When comparing with Eqs.(\ref{c1}, \ref{eq:mixing}), one can see that the limit on 
the Yukawa coupling  from universality is
always less stringent than the limit on the Yukawa coupling 
derived from the maximum neutrino mass and mixing
determined by tritium beta decay and reactor neutrino
experiments, respectively.  This latter
limit is shown as the bottom curve in Figure 1.  
Therefore, one does not obtain
a useful constraint from universality when considering only one extra dimension.     

One can perform the same exercise for two extra dimensions which both have
the same size R.  In the perturbative limit, the mass eigenstates are
$m_{\nu_{k,n}}^2 \approx (k^2 + n^2) / R^2$.
In this case the sum ranges over two indices, $k$ and $n$.  An equation
similar
to Eq.(\ref{eq:limone}) may be derived for this case,
\begin{eqnarray} 
0.0022 \simge 6 \times 10^{-25} {\rm MeV}^2
y^2 \left( {10 \,{\rm TeV} \over M_*} \right)^{2}\left( {{\rm mm} \over R} \right)^{2} 
\Biggl[  \nonumber \\
\sum_{k,n = 1}^{k^2 + n^2 = Q_1 R} \left( {P(A_1,m_{\nu_{k,n}}) \over P(A_1,0)}
-{P(A_2,m_{\nu_{k,n}}) \over
P(A_2,0)}  \right) { 1 \over m_{\nu_{k,n}}^2} 
 - \sum_{k, n = Q_1 R}^{k^2 + n^2 = Q_2 R} {P(A_2,m_{\nu_{k,n}})
\over P(A_2,0)} { 1 \over m_{\nu_{k,n}}^2} \Biggr]
\end{eqnarray}
All two dimensional cases considered here, $M_* > 1 {\rm TeV}$,
fulfill the condition $\zeta^2 << 1$.  
We solve the above equation by converting the sums to integrals,
and find that the limit is again insensitive to R.  Figure 1 also
shows the constraint on $M_*$ and $y$ for the two
dimensional case.  It can be seen that greater precision is needed to
probe scales greater 
than 10~TeV, since above this scale the Yukawa coupling is not
constrained to be less than of order 1.

The Standard Model
value of $V_{ud}$ can be obtained from measured $Ft$ values as in \cite{towner}
with the additional input of $G_F$, the weak coupling constant from muon decay.
Using these quantities, the unitarity sum $V_{ud}^2 + V_{us}^2 + V_{ub}^2$
is calculated to be
smaller than one by two standard deviations.  The extra dimensional
scenario would effect both the muon decay measurement as well as
the $Ft$ values from beta decay.  This introduces additional parameters
and model uncertainties, such as the coupling of the muon neutrino
to the same or another Kaluza Klein tower of neutrinos.  If, for
example,  the muon
neutrinos did not couple to any bulk neutrinos, then the calculated
value of $V_{ud}$ should appear too large and an extra dimensional
scenario could not account for the shortfall in the unitarity sum
through the determination of $V_{ud}$ from beta decay.

Next we make contact with another much discussed test of universality 
using charged pion decays \cite{das}. The
pion decays, $\pi \rightarrow e \nu_e$ and $\pi \rightarrow \mu \nu_{\mu}$ can
be probed kinematically.  Since the electron in the decay is relativistic, while
the muon is not, the contributions of a massive neutrino differ considerably
in the two decays; a massive electon neutrino will make a larger difference 
than a massive muon neutrino.  In the case of massless neutrinos the 
standard model ratio is
$\Gamma(\pi \rightarrow e \nu) / \Gamma(\pi \rightarrow \mu \nu) = 1.233 \times 10^{-4}$.
The experimentally measured quantity is $R^{\pi e}_{\pi \mu} = (1.230 \pm
0.004)
\times 10^{-4}$ \cite{pimeasurement}.  Therefore, the extradimensional
contribution to these decays can not exceed 0.3\%.  The contribution has been 
calculated explicitly by  \cite{das} and for the $n=2$ case with two
families of bulk neutrinos. The
contribution is approximately given by
\begin{equation}
\label{eq:pi}
7.52 \pi \times 10^2 y_e^2 \left({1 {\rm TeV} \over M_*} \right)^2 - 6.23 \pi \times 10^{-1} 
\left( {1 {\rm TeV} \over M_*} \right)^2 y_\mu^2  \simle 0.003.
\end{equation}
With this reaction alone, the only way to obtain a constraint on either $y_e$ or $y_\mu$ 
as a function of $M_*$ is to make an assumption about one of the couplings such as, $y_e \sim 1$ 
or $y_\mu \sim 1$.  Ideally, one would obtain a separate constraint on one
of the
couplings and use the pion decay data to limit the other.  Taking the constraint
on $y_e$ from beta decay universality discussed previously we find that
another two orders of
magnitude in precision in the beta decay data 
would be required in order to probe $y_\mu$. However, a complete
discussion must include an analysis of neutrino oscillations which is
beyond the scope of this paper; however, see \cite{bar} for a discussion.

\section{Kinematic searches}

In addition to changing the apparent $ft$ values, bulk neutrino scenarios
can 
in principle also be probed kinematically.  Here we demonstrate the consequences of
the extra dimensional scenario on nuclear recoil momenta, and electron spectra.
Massive neutrino searches in beta decay have focused on examining beta spectra
\cite{nue, tritium, copper, fluorine}  and on examining the
nuclear recoil spectra \cite{recoil, argon}.  
A tower of Kaluza-Klein bulk states will
produce a somewhat different signal than just one massive neutrino.  We examine 
the signature in both of these types of spectra from the KK tower of states.
Larger mixing scenarios may be probed by tritium beta decay searches for eV
range
neutrinos.  However, as we shall see below,the model of extra dimensions
cannot currently be
constrained by searches for heavy (keV-MeV)  
neutrinos.

With a massive electron neutrino, the number of electron neutrinos
as a function of momentum would be,
\begin{equation}
{N(p_e) \over p_e^2 F(Z,p_e)} = {1 \over m_e^5} {1 \over ft} \left[ (Q - E_e + m_e)^2 - m_\nu^2 \right]^{1/2} (Q - E_e + m_e)
\end{equation}
We show the normalized Kurie plot, $(N(p_e)/(p_e^2 F(Z,p_e)))^{1/2}$ vs. $E_e$
for $^{38m}{\rm K}$ at the top of Figure \ref{fig:kurie}, for the case of
$m_\nu = 0$.  The Q- value for this decay is 5.02 MeV.

In the case of one extra dimension, the function is modified to
\begin{equation}
{N(p_e) \over p_e^2 F(Z,p_e)}  = {1 \over m_e^5} {1 \over ft_{xd}} 
\sum_{k=0}^{k_{max}} |U_{ke}|^2  \left[ (Q - E_e + m_e)^2 - m_{\nu_k}^2
\right]^{1/2} (Q - E_e + m_e).
\end{equation}
Here $|U_{ke}|^2$ is the mixing of the kth mass eigenstate with the electron
neutrino.  For heavy 
massive neutrino searches (keV-MeV) very small mixing 
is relevant as discussed below.  However, for massive neutrino searches in
the eV range, such as tritium beta decay, a slightly larger mixing is
applicable.  Several
states will give a different signature than just one or two massive states
in a kinematic study. We
illustrate this in the lower panel of Figure \ref{fig:trit}.  This panel 
shows the beta endpoint spectrum for tritium for three cases.  The
dashed line shows the spectrum for a single massive neutrino of
2.3 eV.  The solid line shows the result for a scenario with one
extra dimensions, and $\zeta=.13$, $1/R = 25$ eV giving $m_D = 2.3 \, {\rm eV}$.
The first ten mass eigenstates were calculated by explicit diagonalization
of the matrix, Eq. (\ref{M2}).  The first three occur at 2.3 eV, 25 eV
and
50 eV, with mixings $|U_{ke}|^2$ of 0.974, 0.017 and 0.004 respectively.
As can be seen from the figure, the Kaluza-Klein tower produces several bumps
as well as less counts slightly away from the 
endpoint than a single massive neutrino.  For comparison 
a two neutrino mixing scenario is shown as the dot-dashed line. The
mass eigenstates are at 2.3 eV and 25 eV with a mixing of $\sin^2 \theta \approx 2.5\%$.

We now turn to searches for massive neutrinos in the keV and MeV range.
In the case of one extra dimension and small mixing, the function is modified to
\begin{eqnarray}
{N(p_e) \over p_e^2 F(Z,p_e)}  = {1 \over m_e^5} {1 \over ft_{xd}} && \Biggl[ \left(
1 - {\pi^2 \zeta^2 \over 6} \right) ( Q - E + m_e) + \nonumber \\
&&
\sum_{k=1}^{k_{max}} {\zeta^2 \over k^2}  \left[ (Q - E_e + m_e)^2 - m_{\nu_k}^2
\right]^{1/2} (Q - E_e + m_e) \Biggr]
\end{eqnarray}
We have normalized to the number of counts in order to eliminate the uncertainty in $ft_{xd}$.  Therefore, we replace  $N(p_e) \rightarrow N(p_e)/N$, where for
one extra dimension, $N = \int N(p_e) dp_e$.
We plot the results for the extra dimensional scenario as the ratio,
\begin{equation}
 {\cal R} =\left[{N(p_e) \over N p^2 F(Z,p)} \right]^{1/2}_{xd} / \left[{N(p_e)
\over N p^2 F(Z,p)} \right]^{1/2}_{SM}.
\end{equation}
Two cases are
shown in Figure \ref{fig:kurie}.  The panel
 in the bottom left hand corner shows a situation where
many modes can contribute to the decay; $R = 10^{-8} {\rm mm}$, $m_D = 1 \, {\rm eV}$.
The mass of the lightest mode is at about 0.02 MeV and has a mixing of
around $\zeta^2 =  5 \times 10^{-9}$.  Nuclei with smaller Q values
will have a slightly larger signature for the continuum case.  

The experimental limits on the
mixing angle for a single 17 keV neutrino, for example \cite{tritium} are
of order $10^{-3}$.  Since the mixing of the  $\sim 20 \, {\rm keV}$ modes from
this extra dimensional scenario are considerably weaker, these modes are not
 likely to be 
detectable with present experimental data.   The mixing
of a $\sim$1 MeV neutrino is smaller still, since the mixing goes as 
$\zeta / k = \sqrt{2} m_D R / k$.  If the lowest KK state occurs around 1 MeV, this implies a
size $R \approx 2 \times 10^{-10} {\rm mm}$ and a mixing of $\zeta^2 \sim 3 \times 10^{-13}$.
The point here is that the mixing of the a high mass KK mode is proportional to 
$m_D$ which is
constrained by tritium beta decay experiments.  Therefore, in this simple
model the mixing of keV and MeV neutrinos will always be small. 
The bottom right hand corner of  Figure \ref{fig:kurie} shows 
contributions from three separate neutrino
masses, 1, 2 and 3 MeV.
The bumps shown in the lower right panel are unobservable,
since current constraints on the mixing of an MeV neutrino are many orders of
magnitude less ($\sin^2 2\theta \sim 10^{-3}$).

We also consider the method of detecting massive neutrinos by measuring
nuclear recoil spectra.  In the presence of a massive neutrino, the nuclear
recoil spectrum for a pure Fermi transition has the shape,
\begin{equation}
P_r(Z,r) = {1 \over 2} \int^{E_{max}}_{E_{min}}  F(Z,E_e) \left[ r E_e E_\nu + r { a \over 2}
(r^2 - p_e^2 -p_{\nu}^2) \right] dE_e
\end{equation}
Here, $r$ is the recoil momentum of the nucleus and $a = 1$.  For a pure
Gamow-Teller
transition, $a = -1/3$. The limits of integration depend on the
mass of the neutrino and are given by
\begin{equation}
E_{max,min} = {1 \over 2} \left[{E_0 (E_0^2 - r^2 + me^2 -m_\nu^2) \pm r \sqrt{(E_0^2 - r^2 -m_e^2 -m_\nu^2)^2 - 4 m_\nu^2 m_e^2} \over E_0^2 - r^2} \right]
\end{equation}
Here $E_0 = Q + m_e$ is the total energy available for the decay. We again normalize the
distribution by integrating over all possible recoil energies and finding $P_r(m_\nu)$,
for each given neutrino
mass.  The distribution for $m_\nu = 0$ is shown on the top panel of Figure
\ref{fig:recoil}.
The normalized probability distribution for the scenario with one extra dimension looks
like,
\begin{equation}
P_r(r)  = {(1 - {\pi^2 \zeta^2 \over 6}) P_r(r,m_\nu \approx 0) + \sum_k {\zeta^2 \over k^2}
P_r(r,m_\nu \approx k/R) \over P_r}
\end{equation}
where the normalization is
\begin{equation}
P_r = (1 - {\pi^2 \zeta^2 \over 6}) \int P_r(r,m_\nu \approx 0) dr +
\sum_k {\zeta^2 \over k^2}
\int P_r(r, m_\nu \approx k/R) dr
\end{equation}

To see the effect of the bulk neutrino states on recoil spectra, we
again plot normalized ratios, $(P_r(r)/P_r)_{xd} / (P_r(r)/P_r)_{SM}$
The results are shown in Figures \ref{fig:recoil} for
the Fermi transition in $^{38m}{\rm K}$.  The top panels
show the normalized spectra for a single zero mass neutrino.
The bottom left panels show the effect when many
neutrino modes can contribute to the decay, while the bottom right
panels show the case where only a few neutrinos can contribute.
However, as in the case
with the Kurie plots, the mixing of the KK modes is several orders
of magnitude too small to be detected with current experimental 
measurements.

\section{Conclusions}
\label{sec:concl}

We have studied the effects of bulk neutrinos in nuclear $\beta$ decays in
a simple model which only allows the SM singlet state to propagate in
one or more compactified extra dimensions. All other SM particles are 
confined to a 3-brane and have no KK excitations. There are many 
well studied nuclei with different Q values that allow us to use 
universality to place constraints on the Yukawa coupling and the
fundamental scale $M_*$. This information is complementary to the
various studies of physics of higher dimensions using the gravitational
force or KK graviton probes.

Our study shows that for $n=1$ the most stringent constraint on the parameters of
the bulk neutrinos arise from the direct measurement of the mass of $\nu_e$
 and from reactor neutrino oscillation experiments. Universality
tests do not give additional information. On the other hand for $n=2$ the universality
test restricts the Yukawa coupling to be less than unity for $M_* \leq 10 \, \rm TeV$. Future
experiments with radioactive beams measuring decays with Q values of order 10 MeV and 
improved accuracy with current measurements that can push this by a factor of
ten will be most welcome. 

The Kurie plot and the recoil nuclear spectrum in some selected decays are
also given. For probes of massive Kaluza-Klein states we find the nuclei
with smaller Q values are more sensitive. Currently, for medium and heavy
nuclei, the sensitivity
of experiments designed for massive neutrino searches 
are several orders of magnitude below what is needed to
probe mixing of the massive KK neutrinos in these reactions. This may be
due to the simplicity of the model we have considered and hopefully
such studies
will prove useful for more realistic models with enhanced mixings that also address the
family problem. We also point out that our study may be used in connection
with other data to constrain parameters of second and third 
generation mixing with bulk neutrinos.
These scenarios are more model dependent and contain more parameters. We give
an illustration of this point for charged pion decays.

On the other hand we find that the tritium $\beta$ decay spectrum can be
significantly altered by the presence of KK bulk neutrino states in the eV
range. They can reveal themselves in the shape of endpoint region of the Kurie
plot if the mixings are favorable.
 
It is useful to compare the results from universality limits to the ones derived from 
astrophysical constraints which are indirect. The latter relies on the
energy loss carried away by bulk neutrinos.
For  example with limits derived from the neutrino magnetic moment \cite{MN}
for $n=2$ there is essentially no constraint on the neutrino magnetic moment
induced from the sort of extra dimensional scenario presented here, while the
limit on the Yukawa coupling is of order 1 for $M_* \sim 10 \, {\rm TeV}$ 
from universality.
For $n=1$ one obtains a constraint from the mu and tau neutrino magnetic moment
whereas  universality gives no useful constraints. As mentioned before in this case for 
electron neutrinos the strongest constraint comes from the electron
neutrino mass limit and oscillation experiments.

\section{Acknowledgements}
\label{sec:ack}
We wish to thank J. Behr and M.Trinczek for helpful discussions. One of us
(J.N.N.) would like to thank Prof. T.K. Lee of the National Center for
theoretical Science for his hospitality.

This research is partly supported by the Natural Science and Engineering
Council of Canada.

\clearpage
\newpage
\begin{figure}
  \caption{\label{fig:limit}
The upper curve
  shows the limit on the extra dimensional Yukawa coupling from nuclear beta
decay as a function of higher dimensional scale $M_*$.  This constraint is
for a number of extra dimensions, n=2.  The lower curve is for n=1. It
shows the limit
on the extra dimensional Yukawa coupling from the tritium beta decay
neutrino mass limit and the CHOOZ oscillation probability 
limit. This lower curve is a tighter constraint than the
universality test for one extra dimension.}
  \end{figure}

\begin{figure}
  \caption{\label{fig:trit}
The top panel shows a the endpoint of a
Kurie plot for tritium beta decay, for which the
Q value is 18.6 keV.  The horizontal axis is electron kinetic energy - Q.  
The bottom panel shows the same region with three different neutrino
mass scenarios.  This plot shows the subtraction of
zero mass neutrino Kurie plots from the
massive neutrino Kurie plots.
The dashed line shows the spectrum for a single mass
neutrino of 2.3 eV.  The solid line shows the spectrum for one
extra dimension with $\zeta =.13$ and $1/R = 25 \, {\rm eV}$. The
dot-dashed line shows a two neutrino mixing scenario with
states at 2.3 eV and 25 eV and a mixing of $2.5\%$}
  \end{figure}

  \begin{figure}
  \caption {\label{fig:kurie}
The top panel shows a normalized Kurie plot for $^{38m}{\rm K}$.
The horizontal axis is total electron energy. The dashed line
corresponds to a zero mass neutrino, while the solid
line corresponds to a neutrino with mass 2.3 eV.
The bottom two panels show ratios of the normalized Kurie plot in
the standard model to the normalized Kurie plot in the extra dimensional
scenario.  The bottom left panel, shows the effect for $R = 10^{-8} \, {\rm mm}$
and a Dirac mass term of $ m_D = 1 \, {\rm eV} $.  The bottom right panel was produced 
with the
parameters $R = 2 \times 10^{-10} \, {\rm mm}$ and $ m_D = 1 \, {\rm eV}$.}
 \end{figure}

\begin{figure}
  \caption{\label{fig:recoil}
The top panel shows a normalized nuclear recoil spectrum for $^{38m}{\rm K}$.
The bottom two panels show ratios of the normalized spectrum in
the standard model to the normalized spectrum in the extra dimensional
scenario.  The bottom left panel, shows the effect for $R = 10^{-8} \, {\rm mm}$ 
and a Dirac mass term of $ m_D = 1 \, {\rm eV} $.  The bottom right panel was produced
 with the
parameters $R = 2 \times 10^{-10} \, {\rm mm}$ and $ m_D = 1 \, {\rm eV}$.}
 \end{figure}

\begin{figure}
\epsfxsize=14cm
\centerline{\epsffile{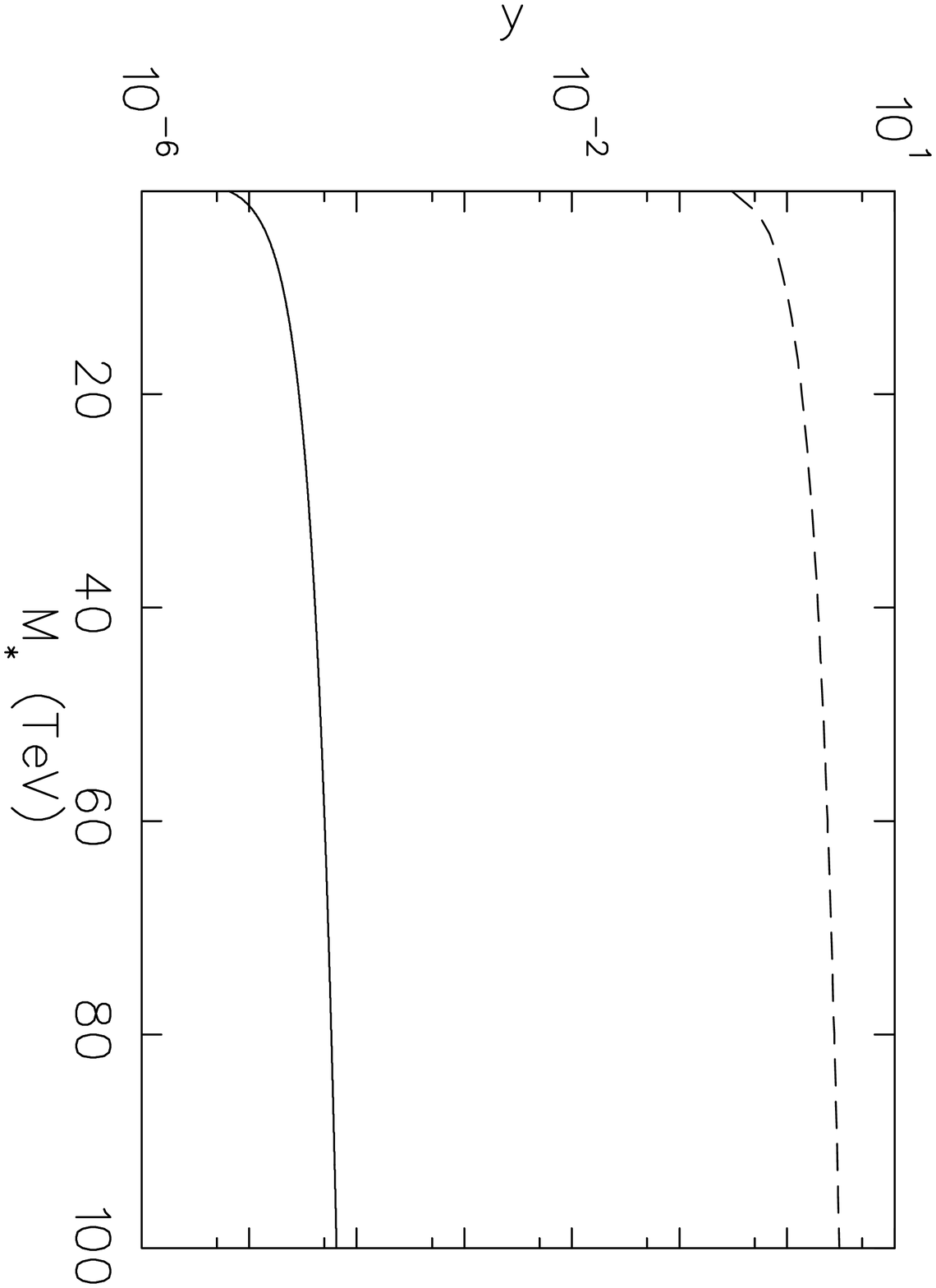}}
\end{figure}
\begin{figure}
\epsfxsize=14cm
\centerline{\epsffile{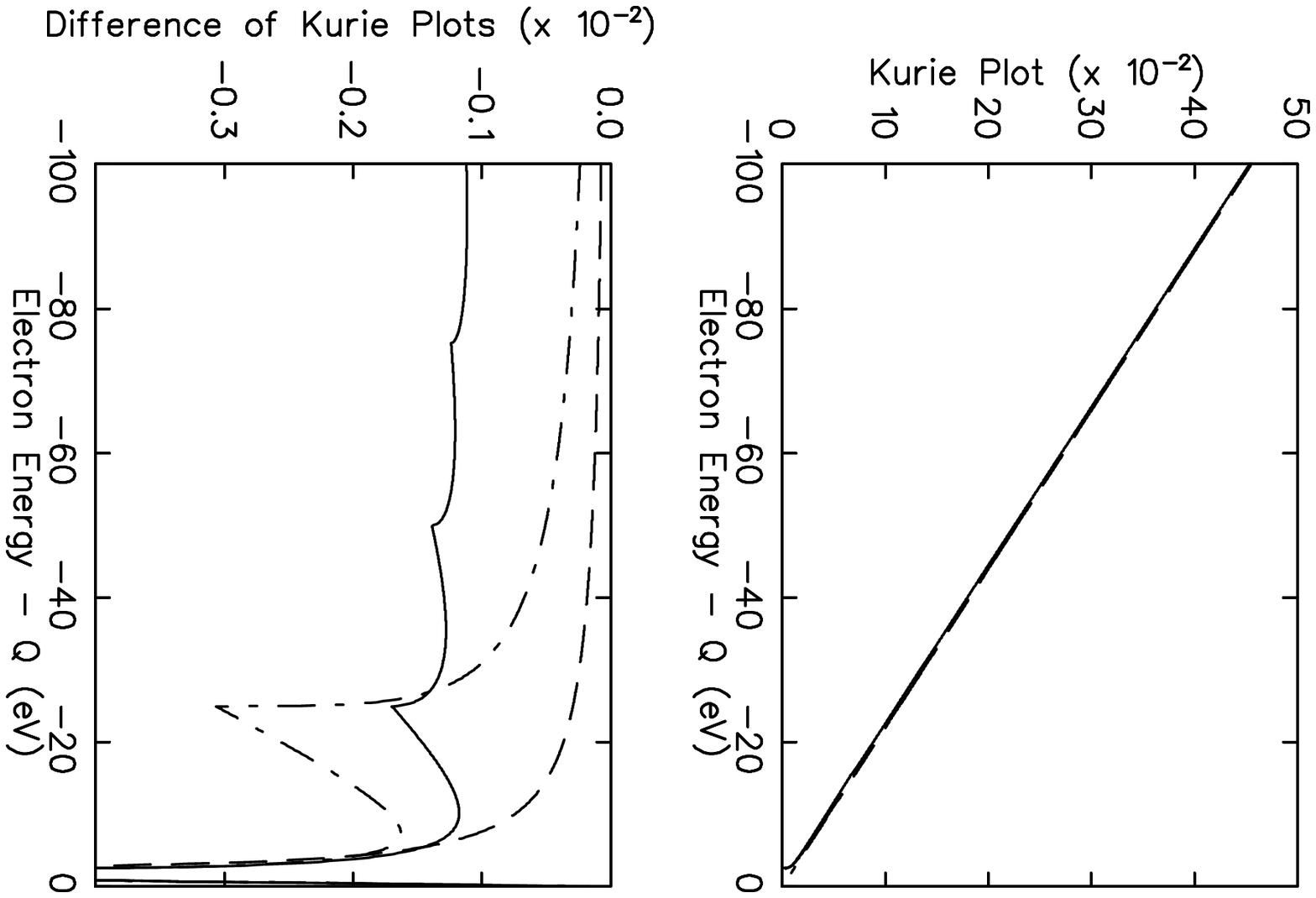}}
\end{figure}
\begin{figure}
\epsfxsize=14cm
\centerline{\epsffile{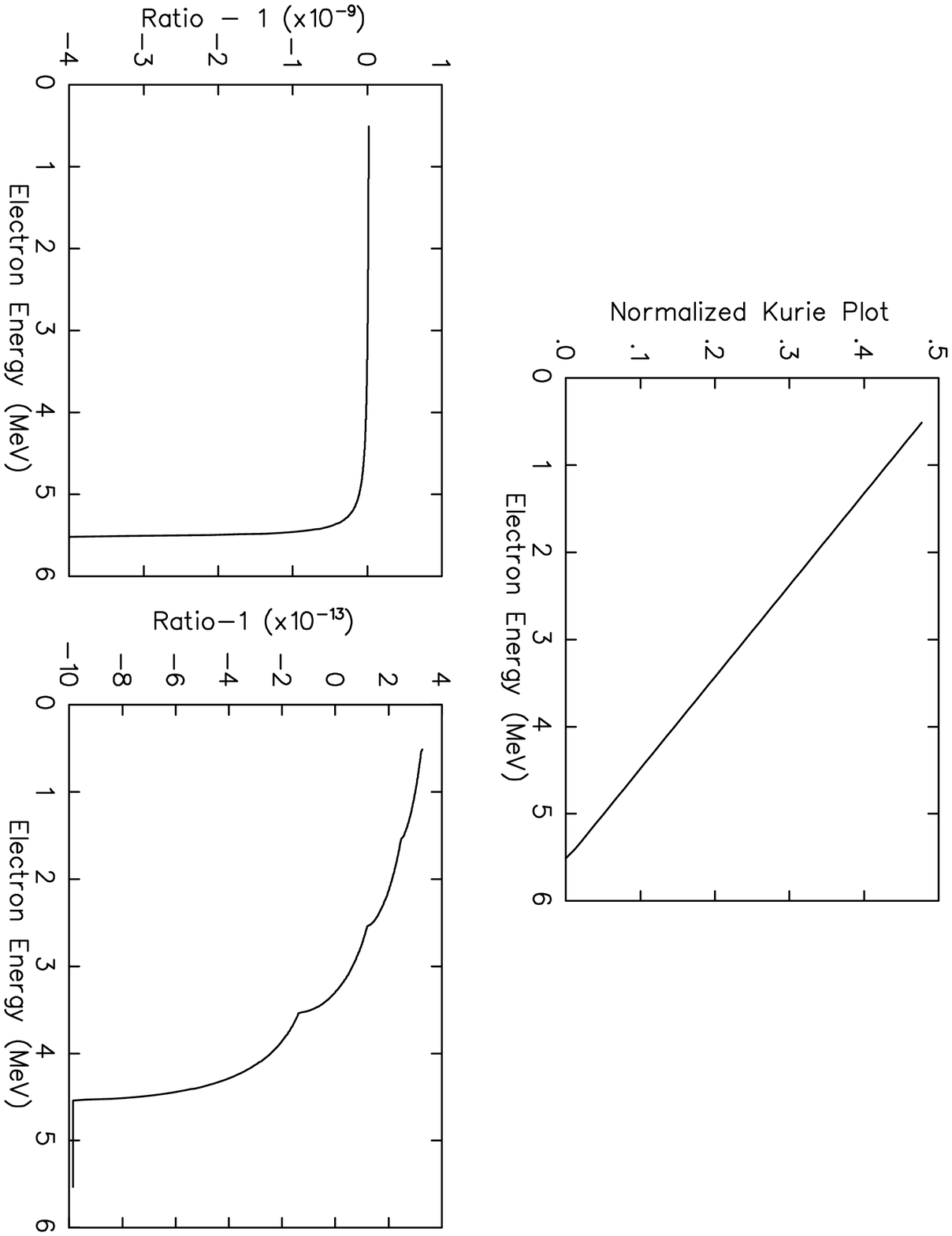}}
\end{figure}
\begin{figure}
\epsfxsize=14cm
\centerline{\epsffile{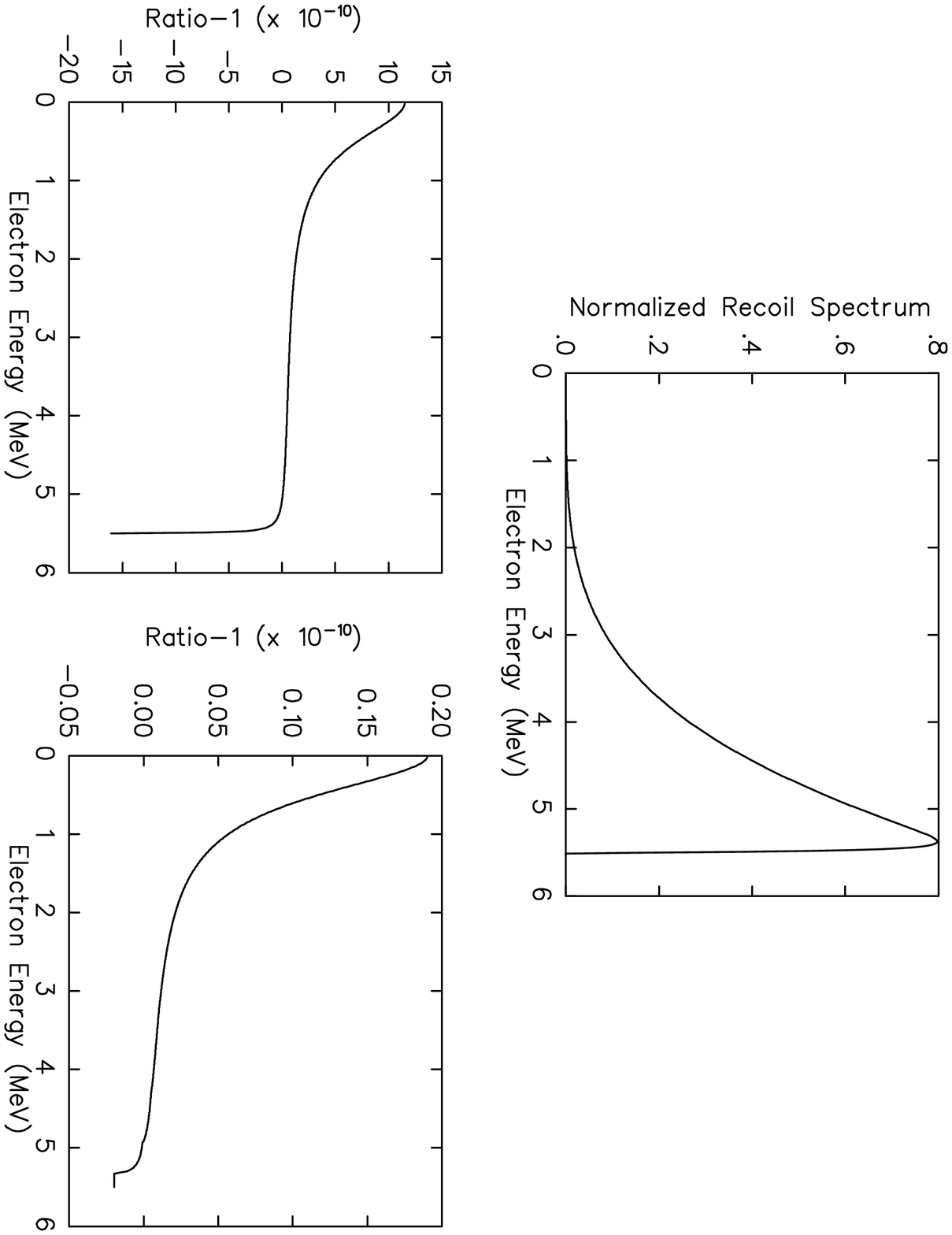}}
\end{figure}

\end{document}